\documentclass[12pt,english]{article}
\usepackage[latin9]{inputenc}
\usepackage{mathtools}
\usepackage{amsmath}
\usepackage{amssymb}
\usepackage{babel}



\begin{document}
\title{On the analog of the Kolmogorov-Arnold superposition representation
for continuous functions of several $p$-adic variables}
\author{A.\,P.~Zubarev \\
 \textit{ Physics Department, Samara University, } \\
 \textit{ Moskovskoe shosse 34, 443123, Samara, Russia} \\
 \textit{Natural Science Department, } \\
 \textit{Samara State University of Railway Transport,} \\
 \textit{Perviy Bezimyaniy pereulok 18, 443066, Samara, Russia} \\
 e-mail:\:\texttt{apzubarev@mail.ru} }
\maketitle
\begin{abstract}
It is shown that any continuous function depending on several $p$-adic
variables, each of which is defined on the ring of $p$-adic integers
$\mathbb{Z}_{p}$, can be represented as a superposition of continuous
functions of one $p$-adic variable. This statement is true for both
functions with values in $\mathbb{R}$ and functions with values in
$\mathbb{Q}_{p}$.
\end{abstract}

\section{Introduction}

Any continuous real-valued function of several real variables can
be represented as a superposition of continuous functions of one real
variable. This statement is stated by the well-known Kolmogorov-Arnold
Superposition Theorem \cite{Kol,Arnold}, which in one of the modern
formulations \cite{Spr,Lor,Hed,Kah} is as follows. For each $n\geq2$
there exist continuous functions $\phi_{k}:\:I=\left[0,1\right]\subset\mathbb{R}$,
$k=1,2,\ldots,2n+1$ and constants $\lambda_{i}$, $i=1,2,\ldots,n$
such that the following holds true: for each continuous function $f:\:I^{n}\rightarrow\mathbb{R}$
there exists a continuous function $g:\:I\rightarrow\mathbb{R}$ such
that
\begin{equation}
f\left(x_{1},x_{2},\ldots,x_{n}\right)=\sum_{k=1}^{2n+1}g\left(\sum_{i=1}^{n}\lambda_{i}\phi_{k}\left(x_{n}\right)\right),\label{K_A}
\end{equation}
moreover the functions $\phi_{k}$ and the constants $\lambda_{i}$
are independent from the represented function $f$ and do only depend
on the dimension $n$.

Recently, the Kolmogorov-Arnold theorem has been used in the process
of creating new generation neural networks (so-called Kolmogorov-Arnold
networks), in which the modeling of complex functions is required.
One of the first architectures of such a network was presented in
\cite{Hech}, and it has demonstrated its high efficiency for function
approximation problems. At present, a great number of works are devoted
to the development and application of Kolmogorov-Arnold networks (see,
for example, \cite{Sch,Liu} and references therein).

In this paper we pose the question: what will the representation (\ref{K_A})
look like for the case when $f\left(x_{1},x_{2},\ldots,x_{n}\right)$
is a continuous function depending not on real but on $p$-adic variables?
We consider cases where the function $f$ is both real-valued and
$p$-adic-valued. It turns out that in these cases there also exists
a representation similar to (\ref{K_A}), but which is much simpler.
Moreover, obtaining such a representation does not require deep mathematical
considerations and its derivation is quite transparent. However, since
it is not presented in the literature on $p$-adic mathematical physics
known to us (see \cite{ALL,ALL_1} for a review), we considered it
necessary to draw attention to it.

Let us first present the minimal information from $p$-adic analysis
(see, e.g., \cite{VVZ,Katok,KS}) that is necessary to understand
this paper. Let $\mathbb{Q}$ be a field of rational numbers and let
$p$ be a fixed prime number. Any rational number $x\neq0$ is representable
as $x=p^{-\gamma}\dfrac{a}{b}$, where $a$, $\gamma$ are integers,
$b$ are natural number and $a$ and $b$ are not divisible by $p$
and have no common multipliers. $p$-Adic norm of the number $x\in\mathbb{Q}$
is defined by the equality $\left|x\right|_{p}=p^{\gamma}$, $\left|0\right|_{p}=0$.
The completion of the field of rational numbers by the $p$-adic norm
forms the field of $p$-adic numbers $\mathbb{Q}_{p}$. The metric
$d\left(x,y\right)=\left|x-y\right|_{p}$ turns $\mathbb{Q}_{p}$
into a complete separable, totally disconnected, locally compact ultrametric
space. In the literature, the following notations are commonly used
for subsets of the field $\mathbb{Q}_{p}$: $B_{r}\left(a\right)=\left\{ x\in\mathbb{Q}_{p}\vcentcolon\left|x-a\right|_{p}\leq p^{r}\right\} $
is a ball of radius $p^{r}$ centered at $a$ (any point belonging
to the ball can be chosen as its center); $S_{r}\left(a\right)=\left\{ x\in\mathbb{Q}_{p}\vcentcolon\left|x-a\right|_{p}=p^{r}\right\} $
is a sphere of radius $p^{r}$ centered at $a$; $B_{r}\equiv B_{r}\left(0\right)$
is a ball of radius $p^{r}$ centered at $0$; $\mathbb{Z}_{p}\equiv B_{0}\left(0\right)$
is the ring of $p$-adic integers. The canonical representation of
a $p$-adic number $x\in\mathbb{Q}_{p}$ with the norm $\left|x\right|_{p}=p^{\gamma}$
is
\begin{equation}
x=p^{-\gamma}\sum_{i=0}^{\infty}x_{i}p^{i},\label{p_adic_ser}
\end{equation}
where $0\leq x_{i}\leq p-1$ for $i>0$ and $0<x_{0}\leq p-1$. The
$n$-dimensional space of $p$-adic numbers $\mathbb{Q}_{p}^{n}$
consists of points $x=\left(x_{1},x_{2},\ldots,x_{n}\right)$, where
$x_{i}\in\mathbb{Q}_{p}$, $i=1,2,\ldots,n$. The $p$-adic norm on
$\mathbb{Q}_{p}^{n}$ is defined as $\left|x\right|_{p}=\max_{i}\left(\left|x_{i}\right|_{p}\right)$.
This norm generates an ultrametric distance $d\left(x^{(1)},x^{(2)}\right)$
on $\mathbb{Q}_{p}^{n}$ for any two points $x^{(1)},x^{(2)}\in\mathbb{Q}_{p}^{n}$:
$d\left(x^{(1)},x^{(2)}\right)=\max_{i}\left(\left|x_{i}^{(1)}-x_{i}^{(2)}\right|_{p}\right)$.
The space $\mathbb{Q}_{p}^{n}$ (like $\mathbb{Q}_{p}$) is a complete
locally compact and totally disconnected ultrametric space. The space
$\mathbb{Z}_{p}^{n}$ is the set of points $x=\left(x_{1},x_{2},\ldots,x_{n}\right)$,
where $x_{i}\in\mathbb{Z}_{p}$, $i=1,2,\ldots,n$. A real-valued
function $f\left(x\right)$ on $\mathbb{Q}_{p}^{n}$ (on $\mathbb{Z}_{p}^{n}$)
is a mapping $\mathbb{Q}_{p}^{n}\rightarrow\mathbb{R}$ ($\mathbb{Z}_{p}^{n}\rightarrow\mathbb{R}$).
Similarly, a $p$-adic-valued function $f\left(x\right)$ on $\mathbb{Q}_{p}^{n}$
( on $\mathbb{Z}_{p}^{n}$ ) is a map $\mathbb{Q}_{p}^{n}\rightarrow\mathbb{Q}_{p}$
($\mathbb{Z}_{p}^{n}\rightarrow\mathbb{Q}_{p}$). As in traditional
analysis, a function $f\left(x\right)$ is said to be continuous on
$\mathbb{Q}_{p}^{n}$ (on $\mathbb{Z}_{p}^{n}$ ) if for all $x,\:y\in\mathbb{Q}_{p}^{n}$
(for all $x,\:y\in\mathbb{Z}_{p}^{n}$) we have $\lim_{d\left(x,y\right)\rightarrow0}\left|f\left(x\right)-f\left(y\right)\right|=0$
for a real-valued function and $\lim_{d\left(x,y\right)\rightarrow0}\left|f\left(x\right)-f\left(y\right)\right|_{p}=0$
for a $p$-adic-valued function.

In Sections 2 and 3, we formulate and prove two simple theorems that
are the result of this paper.

\section{The superposition theorem for real-valued functions}

\textbf{Theorem 1.}Let the function $f\left(x_{1},x_{2},\ldots,x_{n}\right):\:\mathbb{Z}_{p}^{n}\rightarrow\mathbb{R}$
be continuous on $\mathbb{Z}_{p}^{n}$. Then it can be represented
as a superposition

\begin{equation}
f\left(x_{1},x_{2},\ldots,x_{n}\right)=g\left(\sum_{k=1}^{n}\left(n\left(p-1\right)+1\right)^{-k+1}\phi_{n}\left(x_{k}\right)\right),\label{main}
\end{equation}
where $g$ is a continuous function $\mathbb{R}\rightarrow\mathbb{R}$,
and $\phi_{n}$ is a continuous mapping $\mathbb{Z}_{p}\rightarrow I$
of the form

\begin{equation}
\phi_{n}\left(\sum_{j=0}^{\infty}x_{j}p^{j}\right)=\sum_{j=0}^{\infty}x_{j}\left(n\left(p-1\right)+1\right)^{-nj-1}.\label{phi_n}
\end{equation}

\textbf{Proof of Theorem 1.} Let $x=\sum_{j=0}^{\infty}x_{j}p^{j}\in\mathbb{Z}_{p}$ and $\varphi\left(x\right)$
be a continuous mapping $\mathbb{Z}_{p}\rightarrow I\equiv\left[0,1\right]$
of the form

\begin{equation}
\varphi\left(x\right)=\sum_{j=0}^{\infty}x_{j}\left(n\left(p-1\right)+1\right)^{-j-1}.\label{varphi}
\end{equation}
The image of the mapping $\varphi\left(x\right)$ is the Cantor-like
set $C_{p}\subset I$, which is uncountable, closed, and does not
contain isolated points, and the mapping $\varphi\left(x\right)$
is a homeomorphism \cite{VVZ,Katok,KS}. We will denote the inverse
mapping to (\ref{varphi}) as $x=\psi\left(\tilde{x}\right),\;x\in\mathbb{Z}_{p},\;\tilde{x}\in C_{p}$.

Let $\tilde{x}=\sum_{j=0}^{\infty}x_{j}\left(n\left(p-1\right)+1\right)^{-j-1}\in C_{p}$,
$x_{j}=0,1,\ldots p-1$. Let us define an injective continuous mapping
$\varPhi:\:C_{p}\rightarrow C_{p}$ of the form

\begin{equation}
\varPhi\left(\tilde{x}\right)=\sum_{j=0}^{\infty}x_{j}\left(n\left(p-1\right)+1\right)^{-nj-1}.\label{Phi_s}
\end{equation}
Note that the composition $\phi_{n}\equiv\varPhi\circ\varphi$ gives
the mapping

\begin{equation}
\phi_{n}\left(x\right)=\sum_{j=0}^{\infty}x_{j}\left(n\left(p-1\right)+1\right)^{-nj-1},\label{phi_n_def}
\end{equation}
which is exactly the function appearing in the statement of the theorem.

Next, for a set of elements $\tilde{x}_{1},\dots,\tilde{x}_{n}\in C_{p}$,
where $\tilde{x}_{m}=\sum_{j=0}^{\infty}x_{m,j}\left(n\left(p-1\right)+1\right)^{-j-1}$,
we define the mapping $\varPhi_{n}:\:C_{p}^{n}\rightarrow C_{p}$
as
\begin{equation}
\varPhi_{n}\left(\tilde{x}_{1},\dots,\tilde{x}_{n}\right)=\sum_{k=1}^{n}\left(n\left(p-1\right)+1\right)^{-k+1}\varPhi\left(\tilde{x}_{k}\right)=\sum_{k=1}^{n}\sum_{j=0}^{\infty}x_{k,j}\left(n\left(p-1\right)+1\right)^{-(nj+k-1)-1}=\tilde{z}\in C_{p}.\label{Phi_n}
\end{equation}
Since the exponent $M=nj+k-1$ runs through all non-negative integers exactly once as $j\geq0$ and $1\leq k\leq n$, the digits of all variables are uniquely interleaved without overlaps.

Also for $\tilde{z}=\sum_{m=0}^{\infty}z_{m}\left(n\left(p-1\right)+1\right)^{-m-1}\in C_{p}$
we define the mapping $\varPsi_{n}:\:C_{p}\rightarrow C_{p}^{n}$
as

\begin{equation}
\varPsi_{n}\left(\tilde{z}\right)=\left(\chi_{1}\left(\tilde{z}\right),\dots,\chi_{n}\left(\tilde{z}\right)\right)\equiv\left(\tilde{x}_{1},\dots,\tilde{x}_{n}\right)\text{,}\label{Psi_n}
\end{equation}
with functions

\begin{equation}
\chi_{k}\left(\tilde{z}\right)=\sum_{j=0}^{\infty}z_{nj+k-1}\left(n\left(p-1\right)+1\right)^{-j-1}, \quad k=1,\dots,n,\label{chi_s}
\end{equation}
that are continuous on $C_{p}$.

\textbf{Lemma 1.} The mappings (\ref{Phi_n}) and (\ref{Psi_n}) are
mutually inverse and homeomorphic.

\textbf{Proof of Lemma 1. }From the structure of the mappings (\ref{Phi_n})
and (\ref{Psi_n}) it follows that they are mutually inverse and one-to-one.
Let us prove the continuity of the mapping (\ref{Phi_n}). Consider
two points $\mathbf{\tilde{x}}^{(1)}, \mathbf{\tilde{x}}^{(2)}$
from $C_{p}^{n}$. Let $N_k$ be the first position where the expansions of $\tilde{x}_k^{(1)}$ and $\tilde{x}_k^{(2)}$ differ.
Then the square of the Euclidean distance satisfies
\[
d^{2}\left(\mathbf{\tilde{x}}^{(1)},\mathbf{\tilde{x}}^{(2)}\right) \geq c_1 \sum_{k=1}^n \left(n\left(p-1\right)+1\right)^{-2N_k}
\]
for some constant $c_1>0$. The Euclidean distance between the images is bounded above by
\[
\left|\varPhi_{n}\left(\mathbf{\tilde{x}}^{(1)}\right)-\varPhi_{n}\left(\mathbf{\tilde{x}}^{(2)}\right)\right| \leq c_2 \sum_{k=1}^n \left(n\left(p-1\right)+1\right)^{-2N_k}
\]
for some constant $c_2>0$, since the digits are placed at positions $nj+k-1$.
Thus, we have
\[
\lim_{d\left(\mathbf{\tilde{x}}^{(1)},\mathbf{\tilde{x}}^{(2)}\right)\rightarrow0}\left|\varPhi_{n}\left(\mathbf{\tilde{x}}^{(1)}\right)-\varPhi_{n}\left(\mathbf{\tilde{x}}^{(2)}\right)\right|=0
\]
and the mapping (\ref{Phi_n}) is continuous. The continuity of the
mapping (\ref{Psi_n}) can be shown similarly. Lemma 1 is proved.

Consider a continuous function $f\left(x_{1},\dots,x_{n}\right):\:\mathbb{Z}_{p}^{n}\rightarrow\mathbb{R}$.
Next, we consider a sequence of homeomorphic mappings

\begin{equation}
\mathbb{Z}_{p}^{n}\overset{\varphi\times\dots\times\varphi}{\longrightarrow}C_{p}^{n}\overset{\varPhi_{n}}{\longrightarrow}C_{p}\overset{\varPsi_{n}}{\longrightarrow}C_{p}^{n}\overset{\psi\times\dots\times\psi}{\longrightarrow}\mathbb{Z}_{p}^{n}\label{chain}
\end{equation}
By making transformations in the arguments of the function $f$
from the first part of the sequence (\ref{chain}),
we obtain the function
\[
f\left(x_{1},\dots,x_{n}\right)=f\left(\psi\left(\chi_{1}\left(\tilde{z}\right)\right),\dots,\psi\left(\chi_{n}\left(\tilde{z}\right)\right)\right)\equiv\tilde{g}\left(\tilde{z}\right).
\]
Since the functions (\ref{chi_s}) are continuous on $C_{p}$, the
function $\tilde{g}\left(\tilde{z}\right)$ is also continuous on
$C_{p}$.

The function $\tilde{g}$ is defined on the set $C_{p}$, which is
closed. We extend this function to a function $g:\:\mathbb{R}\rightarrow\mathbb{R}$
using the Tietze extension theorem (see, for example, \cite{Dugundji,Ossa,Munkres,Cirstea}).

Next, using the definition of $\varPhi_{n}$ from (\ref{Phi_n}) and the relation
$\varPhi(\varphi(x)) = \phi_{n}(x)$ from (\ref{phi_n_def}), we obtain
\[
g\left(\tilde{z}\right)=g\left(\sum_{k=1}^{n}\left(n\left(p-1\right)+1\right)^{-k+1}\varPhi\left(\varphi\left(x_{k}\right)\right)\right)=g\left(\sum_{k=1}^{n}\left(n\left(p-1\right)+1\right)^{-k+1}\phi_{n}\left(x_{k}\right)\right).
\]
Thus, we get the representation stated in the theorem,
which completes the proof.

\section{The superposition theorem for $p$-adic-valued functions}

\textbf{Theorem 2.} Let the function $f(x_{1},x_{2},\ldots,x_{n}):\mathbb{Z}_{p}^{n}\rightarrow\mathbb{Q}_{p}$
be continuous on $\mathbb{Z}_{p}^{n}$. Then it can be represented
as a superposition
\begin{equation}
f(x_{1},x_{2},\ldots,x_{n})=h\left(\sum_{k=0}^{n-1}p^{k}\omega(x_{k+1})\right),\label{main_2}
\end{equation}
where $h$ is a continuous function $\mathbb{Z}_{p}\rightarrow\mathbb{Q}_{p}$
and $\omega(x)$ is a continuous mapping $\mathbb{Z}_{p}\rightarrow\mathbb{Z}_{p}$
of the form
\begin{equation}
\omega(x)=\sum_{j=0}^{\infty}x_{j}p^{nj}, \quad \text{for } x=\sum_{j=0}^{\infty}x_{j}p^{j}\in\mathbb{Z}_{p}.\label{omega}
\end{equation}

\textbf{Proof of Theorem 2.} Define the mapping $\Phi:\mathbb{Z}_{p}^{n}\rightarrow\mathbb{Z}_{p}$ by
\begin{equation}
\Phi(x_{1},\dots,x_{n})=\sum_{k=0}^{n-1}p^{k}\omega(x_{k+1}).\label{Phi_n_def}
\end{equation}
Substituting the definition of $\omega$, we obtain:
\[
\Phi(x_{1},\dots,x_{n})=\sum_{k=0}^{n-1}\sum_{j=0}^{\infty}x_{k+1,j}p^{nj+k}.
\]
Since the exponent $M=nj+k$ runs through all non-negative integers exactly once, the digits of all variables are uniquely interleaved. No carries occur between blocks corresponding to different $k$.

\textbf{Lemma 2.} The mapping $\Phi:\mathbb{Z}_{p}^{n}\rightarrow\mathbb{Z}_{p}$ defined by (\ref{Phi_n_def}) is a homeomorphism.

\textbf{Proof of Lemma 2.} Bijectivity follows directly from the unique interleaving of digits. The inverse mapping $\Psi:\mathbb{Z}_{p}\rightarrow\mathbb{Z}_{p}^{n}$ explicitly extracts the digits at positions congruent to $k \pmod n$:
\[
\Psi(z)=(\sigma_{0}(z),\dots,\sigma_{n-1}(z)), \quad \text{where } \sigma_{k}(z)=\sum_{j=0}^{\infty}z_{nj+k}p^{j}.
\]
To prove continuity of $\Phi$, consider two points $\mathbf{x}, \mathbf{x}' \in \mathbb{Z}_p^n$. Let $N_k$ be the first position where $x_{k+1}$ and $x'_{k+1}$ differ. The images differ first at position $M^* = n N_{k^*} + k^*$, yielding the estimate
\[
|\Phi(\mathbf{x})-\Phi(\mathbf{x}')|_p \leq d(\mathbf{x},\mathbf{x}')^n.
\]
Similarly, for the inverse mapping $\Psi$, a difference at position $M$ affects only the component with index $k=M \bmod n$ at position $\lfloor M/n \rfloor$, giving
\[
d(\Psi(z),\Psi(z')) \leq p^{(n-1)/n} |z-z'|_p^{1/n}.
\]
Thus, $\Phi$ is a homeomorphism. Lemma 2 is proved.

Now consider a continuous function $f:\mathbb{Z}_{p}^{n}\rightarrow\mathbb{Q}_{p}$. Define $h:\mathbb{Z}_p \to \mathbb{Q}_p$ by
$h(z) = f(\Psi(z))$.
Since $\Psi$ and $f$ are continuous, $h$ is continuous. Substituting $z = \Phi(x_1, \dots, x_n)$, we obtain
\[
f(x_1, \dots, x_n) = h\left(\sum_{k=0}^{n-1}p^{k}\omega(x_{k+1})\right),
\]
which completes the proof of Theorem 2.

\section{Concluding remarks}

We have shown that any continuous function depending on several $p$-adic
variables, each of which is defined on $\mathbb{Z}_{p}$, can be represented
as a superposition of continuous functions of one $p$-adic variable.
This statement is true for both real-valued functions and $p$-adic-valued
functions. Note that this result, formulated in Theorems 1 and 2,
is a rather simple consequence of the uniqueness of the representation
of a $p$-adic number in the form of a canonical decomposition in
the form (\ref{p_adic_ser}). It is precisely because of this uniqueness
that there is a homeomorphism between $B_{r}^{n}=\underbrace{B_{r}\times B_{r}\times\cdots\times B_{r}}_{n}$
and $B_{r}$ of the form $B_{r}\ni x\longleftrightarrow\left(x^{(1)},x^{(2)},\ldots,x^{(n)}\right)\in B_{r}^{n}$,
where $x=p^{-r}\sum_{i=0}^{\infty}x_{i}p^{i}$, $x^{(j)}=p^{-r}\sum_{i=0}^{\infty}x_{i\cdot j}p^{i}$.
Thus, any continuous function on $B_{r}^{n}$ with values in both
$\mathbb{R}$ and $\mathbb{Q}_{p}$ can be uniquely defined by defining
it on $B_{r}$. Since $B_{r}$ is homeomorphic to $\mathbb{Z}_{p}$,
the structure of this correspondence is also determined by superpositions
of the forms (\ref{main}) or (\ref{main_2}).

\section*{Acknowledgments}

The author thanks J. Furno for valuable comments that helped clarify the injectivity conditions in the proof of Theorem 1.

The study was supported by the Ministry of Higher Education and Science
of Russia by the State assignment to educational and research institutions
under project no. FSSS-2023-0009.

\section*{Data Availability Statement}

The data supporting the findings of this study are available within
the article and its supplementary material. All other relevant source
data are available from the corresponding author upon reasonable request.

\end{document}